\documentclass{emulateapj}
\usepackage{apjfonts,psfig}

\begin{document}

\title{Short-duration gamma-ray bursts from off-axis collapsars}
\shorttitle{Short GRBs from collapsars}

\author{Davide Lazzati\altaffilmark{1}, Brian J.
Morsony\altaffilmark{2} and Mitchell C. Begelman\altaffilmark{3,4}}
\email{davide\_lazzati@ncsu.edu}
\shortauthors{Lazzati et al.}

\altaffiltext{1}{Department of Physics, NC State University, 2401
Stinson Drive, Raleigh, NC 27695-8202} 
\altaffiltext{2}{Department of Astronomy, University of
Wisconsin-Madison, 5534 Sterling Hall, 475 N. Charter Street, Madison WI
53706-1582}
\altaffiltext{3}{JILA, University
of Colorado, 440 UCB, Boulder, CO 80309-0440}
\altaffiltext{4}{University of Colorado, Department of Astrophysical and 
Planetary Sciences, 391 UCB, Boulder, CO 80309-0391}

\begin{abstract} We present 2D high-resolution hydrodynamic simulations
of the relativistic outflows of long-duration gamma-ray burst
progenitors. We analyze the properties of the outflows at wide off-axis
angles, produced by the expansion of the hot cocoon that surrounds the
jet inside the progenitor star. We find that the cocoon emission at wide
angles may have properties similar to those of the subclass of
short-duration gamma-ray bursts with persistent X-ray emission. We
compute the predicted duration distribution, redshift distribution, and
afterglow brightness and we find that they are all in agreement with the
observed properties of short GRBs with persistent emission. We suggest
that a SN component, the properties of the host galaxies, and late
afterglow observations can be used as a crucial test to verify this
model. \end{abstract}


\section{Introduction}

Gamma-ray bursts (GRBs) are divided in two classes based on their
duration and on the hardness of their prompt emission spectra (Mazets et
al. 1981; Kouveliotou et al. 1993). Long-duration bursts (LGRBs) were
the first class for which afterglow observations were possible (Costa et
al. 1997). Afterglow observations have allowed us to study them in great
detail. We now know that long GRBs are associated with the deaths of
massive compact progenitor stars (Woosley 1993; Stanek et al. 2003;
Hjorth et al. 2003), take place in compact star-forming galaxies, and
can be seen up to redshift $z=8.2$ (GRB~090423, Tanvir et al. 2009;
Salvaterra et al. 2009).

Much less is known about the class of short-duration bursts (SGRBs) and
their progenitors. It has been suspected for a long time that SGRB
progenitors are binary systems of two compact objects that merge after
their orbital energy has been radiated, mostly in gravitational waves
(Eichler et al. 1989; Lee \& Ramirez-Ruiz 2007; Ramirez-Ruiz \& Lee
2009). With the advent of HETE2 and Swift, afterglow observations of
SGRBs have become possible (Fox et al. 2005; Gehrels et al. 2005; Hjorth
et al. 2005; Villasenor et al. 2005). The localization of SGRBs has
confirmed that some are associated with early type galaxies and explode
at relatively large distances from the center of the host galaxy,
confirming their origin as due to the merger of compact objects (Nakar
et al. 2007; Berger 2009). However, not all SGRBs fall within such a
simple scheme. Some short bursts, especially those showing a plateau of
persistent prompt emission in the X-rays (Lazzati et al. 2001;
Villasenor et al. 2005) are associated with small, star forming
galaxies, and explode close to the center of their hosts (Troja et al.
2008). This led to the speculation that short bursts have a more diverse
family of progenitors than long bursts, possibly involving different
compact objects (neutron stars vs. black holes, Belczynski et al. 2006;
Troja et al. 2008) or transient proto-magnetars (Metzger et al. 2008).
Virgili et al. (2009) studied GRB populations based on a new Type I/Type
II classification (see Zhang et al. 2007, 2009) and came to the
conclusion that most SGRBs closely follow star formation and cannot
therefore originate from a traditional merger path. They suggest that a
large fraction of SGRBs may be associated with massive stars, analogously
to LGRBs. A close association between the progenitors of SGRBs and
LGRBs was theoretically explored within the multi-jet scenario (Toma et
al. 2005ab) and in the cannonball scenario (Dado et al. 2009).

In this letter, we show that LGRB progenitors may produce SGRBs at
large off-axis angles. This letter is organized as follows: in \S~2 we
present our numerical setup, in \S~3 we describe the properties of the
relativistic outflow at large off-axis angles, and in \S~4 we discuss
some tests that we have performed to assure that this model
is consistent with observations of SGRB rates and their afterglows. Our
results are summarized and discussed in \S~5.

\section{Numerical model}

We considered a 16 solar-mass Wolf-Rayet progenitor star, with a
pre-explosion radius $R_\star=4\times10^{10}$~cm (Model 16TI, Woosley \&
Heger 2006). A typical LGRB jet was introduced as a
boundary condition at a distance of $10^9$~cm from the center of the
star, with a constant luminosity $L_{\rm{jet}}=5.33\times10^{50}$~erg/s,
an initial opening angle $\theta_0=10^\circ$, an initial Lorentz factor
$\Gamma_0=5$, and a ratio of internal over rest mass energy $\eta_0=80$,
allowing for a maximum Lorentz factor of
$\Gamma_\infty=\Gamma_0\eta_0=400$, attainable in case of complete,
non-dissipative acceleration. The jet and the progenitor star evolution
were computed with the adaptive mesh, special-relativistic hydrodynamic
code FLASH (Fryxell et al. 2000) for a total laboratory time of 50
seconds. The simulation presented here is analogous to the one presented
in Morsony et al. (2007) and Lazzati et al. (2009), but was carried out
at higher temporal and spatial resolution in order to study short
time-scale variability properties of GRBs (see Morsony et al. 2009, in
preparation).

\begin{figure}[!t]
\plotone{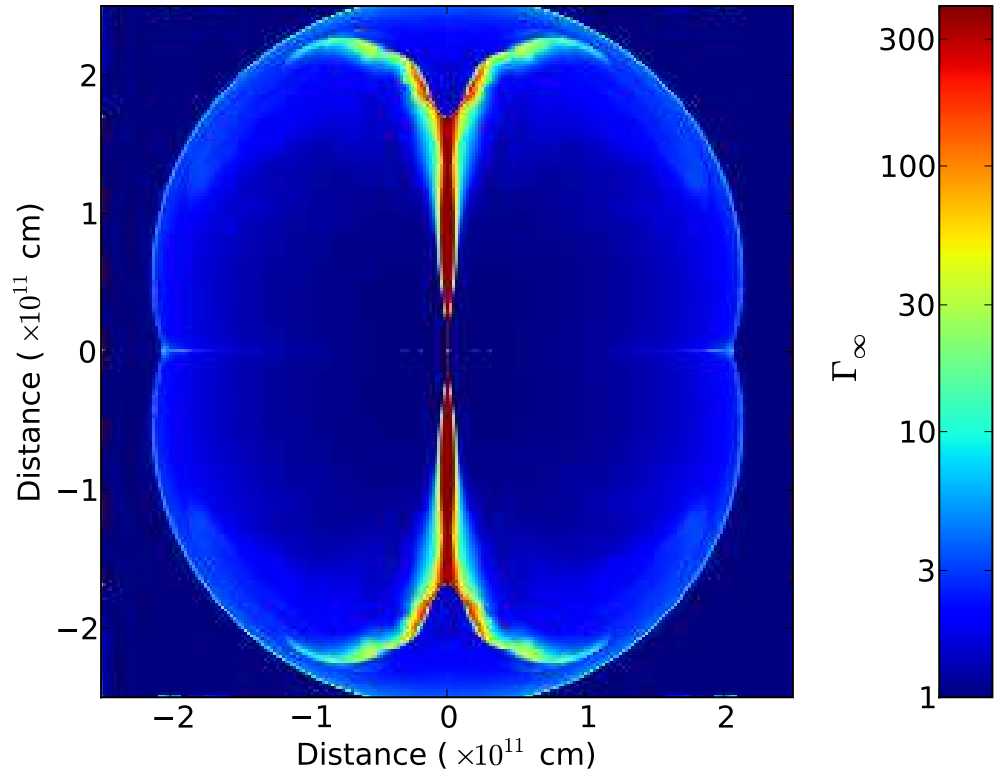}
\caption{{Image of the asymptotic Lorentz factor of the outflow at lab
time $t=13.65$~s from the injection of the jet. The highly relativistic
material of the jet is clearly visible along the axis. Moderately
relativistic material is also seen along the leading edge of an
expanding quasi-isotropic bubble. This is the material that could produce
short GRBs.}
\label{fig:one}}
\end{figure}

We used the results of the simulation to analyze the properties of the
outflow at wide angles, larger than $30\degr$ off-axis.
Figure~\ref{fig:one} shows an image in false colors of the asymptotic
Lorentz factor $\Gamma_\infty$ of the outflow at $t=13.65$~s from the
injection of the jet. The highly relativistic material along the jet
axis is clearly visible in dark red. In addition, a thin shell of
material with an asymptotic Lorentz factor $\Gamma_\infty\sim15$ is seen
surrounding the jet out to very wide angles. In the following, we
discuss the properties and the potential emission of this shell and show
that it can give origin to a SGRB with a persistent tail of low energy
emission.

\section{Off-axis outflow properties}

The propagation of a LGRB jet through its progenitor star is
accomplished at the expense of ``wasting" some of the jet energy that
is recycled into a high pressure cocoon that surrounds the jet
(Ramirez-Ruiz et al. 2002ab, Lazzati \& Begelman 2005). Emission from
the cocoon has been considered as a source for LGRB precursors
(Ramirez-Ruiz et al. 2002a), as an explanation for the steep decay in
the early X-ray afterglow (Pe'er et al. 2006), and as a source of seed
photons for inverse Compton scattering to explain very high energy
components in some LGRBs (Toma et al. 2009). The energy stored in the
cocoon is:
\begin{equation}
E_{\rm{cocoon}}=L_j\left(t_{\rm{br}}-\frac{R_\star}{c}\right)
\simeq7.7\times10^{51}\,L_{j,51}\,R_{\star,11}\,{\rm erg} 
\end{equation}
where $L_{j,51}$ is the jet luminosity in units of $10^{51}$~erg/s,
$R_{\star,11}$ s the progenitor radius in units of $10^{11}$~cm, and
$t_{\rm{br}}$ is the breakout time of the jet on the progenitor surface.
A jet propagation speed of $v_{\rm{head}}\simeq0.3\,c$ has been assumed
to compute the numerical value. The simulation we performed, with
$t_{br}=6.2$~s and $R_\star=4\times10^{10}$~cm, gives
$E_{\rm{cocoon}}=2.6\times10^{51}$~erg and an average
$v_{\rm{head}}=0.22 c$. The cocoon is released approximately at rest on
the progenitor surface and, since it is hot and it has high pressure,
it subsequently accelerates to relativistic speed, maintaining a
constant radial width comparable to the progenitor radius (e.g. Piran
1999). Therefore, should the cocoon be uniform and should it release all
its energy into radiation at constant radius, it would produce emission
for a duration $\delta{t}=\delta{R}/c\sim{R}_\star/c$, approximately
$1$~s for our progenitor star\footnote{Note that this simplified
equation implies that the duration of the transient scales linearly with
the stellar radius. Large stars would therefore produce longer
transients that would not be classified as short GRBs.}.

\begin{figure}[!t]
\plotone{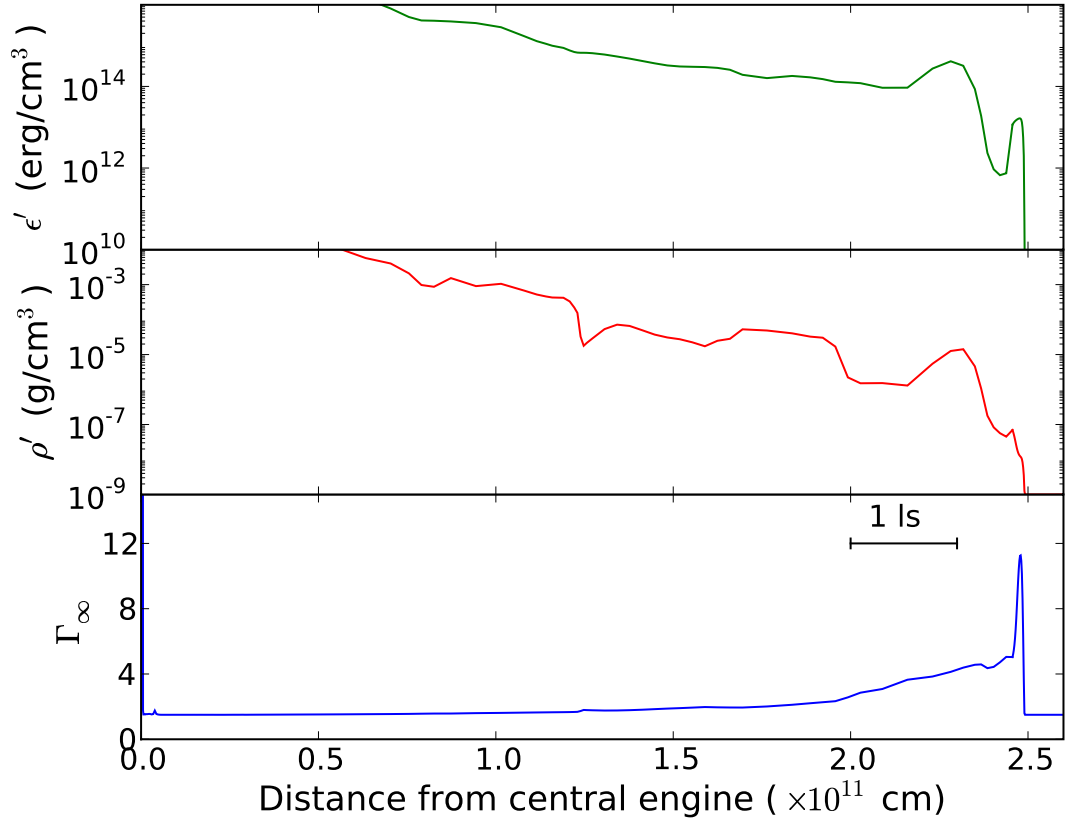}
\caption{{Radial profile of comoving energy density (upper panel),
    comoving baryon density (middle panel) and asymptotic Lorentz
    factor (bottom panel) for an off-axis angle
    $\theta_o=45\degr$. The figure shows that the relativistic
    component of the cocoon is concentrated in a narrow region at the
    leading edge, allowing for a very short emission profile.  The
    distance of a light second is indicated in the bottom panel for
    reference.}
\label{fig:trace}}
\end{figure}

Figure~\ref{fig:trace} shows the radial profile of the cocoon comoving
energy density, comoving baryon density, and asymptotic Lorentz factor
at $\theta_o=45\degr$ and $t=13.65$~s, when the leading edge of the
cocoon has reached a radius $r=2.5\times10^{11}$~cm, the outer edge of
our simulation box. The figure shows that the cocoon develops a radial
structure during its expansion and acceleration outside of the star,
with a leading shell of relativistic material that has a thickness of
only a fraction of a light second. The development of radial
stratification allows for the production of light transients with
duration significantly shorter than the light crossing time of the
progenitor star.

An important feature of the cocoon outflow is that its Lorentz factor is
smaller than the one of the jet material. This is caused by the fact
that the internal energy per baryon in the cocoon is less than the
internal energy per baryon in the jet. The decrease of internal energy
per baryon is due to the combination of two processes. First, part of
the cocoon energy is expended in $pdV$ work for creating the cocoon
cavity inside the star. Second, the material that forms the cocoon is
not purely jet material but is contaminated by mixing with the stellar
material. While numerical simulations can compute very accurately the
$pdV$ work, it is difficult to calculate the amount of mixing, and
therefore the Lorentz factor reported in Figure~\ref{fig:trace} should
be taken with some caution.

\section{Observed properties of short GRBs from collapsars}

We use the results of our high-resolution simulation to compute
observable properties of potential GRBs. First, we compute light-power
curves. These (see also Morsony et al. 2010) are light curves computed
under the assumption that a fraction of the energy in the outflow is
converted into radiation and radiated at a fixed radius. In this
letter, we calculate light-power curves at
$r=2.5\times10^{11}$~cm. This is the largest radius in our
simulation. At an off-axis angle of $45^\circ$, the fireball at this
radius is still optically thick with an optical depth to Thompson
scattering $\tau_T\simeq8$. The flow therefore becomes optically
thin only at a radius $R\sim7\times10^{11}$~cm. Due to the increase in
the shell thickness with distance, our calculations give reliable
light curve properties as long as the emission radius of a real
off-axis GRB is lower than $R\sim2.5\times10^{12}$~cm, beyond which
both the radial spreading of the shell and the effect of the fireball
curvature make the light curve much broader than we calculate. Even
though we do not adopt a physical radiation mechanism, we can separate
high frequency emission from lower frequency emission under the
assumption that the observed peak frequency of the radiation grows
monotonically with the bulk Lorentz factor of the emitting
material. Even though it may seem natural, this assumption has
important consequences. The cocoon shell has a much lower Lorentz
factor than the on-axis jet, but the peak frequencies of LGRBs and
SGRBs are comparable. We therefore assume implicitly that the
radiation mechanism of the cocoon is different from the radiation
mechanism of the jet.

\begin{figure}[!t]
\plotone{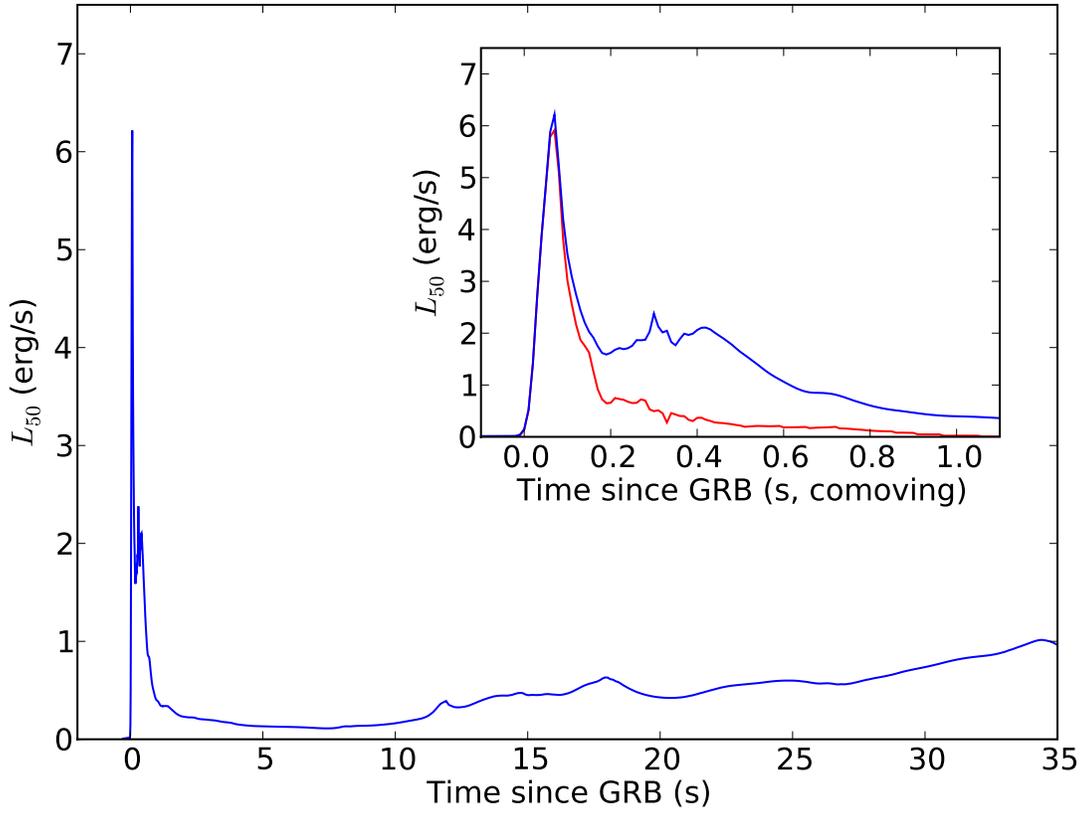} 
\caption{{Bolometric light-power curve of large off-axis angle GRBs
    from collapsars. The curve is computed at a distance
    $R=2.5\times10^{11}$~cm from the GRB engine (the outer edge of the
    simulation box) at an off-axis angle $\theta_o=45\degr$. As a
    consequence of the shell dynamic evolution at or beyond the
    photosphere of the outflow, the real light curve should appear
    slightly broader (see \S~4 for a discussion of the conditions under
    which this light curve is accurate). The main panel shows the
    curve from all the material moving with a minimum bulk Lorentz
    factor $\Gamma\ge2$.  The inset shows a zoom of the initial second
    of the light-power curve, with a red line showing emission from
    material with $\Gamma\ge10$, likely the high energy component of
    the burst.}
\label{fig:lcur}} 
\end{figure}

Figure~\ref{fig:lcur} shows the light-power curve for an off-axis angle
$\theta_o=45\degr$. The emission from material moving with $\Gamma\ge2$
is shown in blue, while the emission from material moving at
$\Gamma\ge10$ is shown in red in the inset. Under the above assumption
for the correlation between photon frequency and bulk Lorentz factor,
the red curve shows high energy emission, while the blue curve shows
also emission components in the X-rays. The light-power curves in the
figure display all the characteristic features of SGRBs with persistent
emission. The initial spike has a duration of $\sim0.1$~s and is
followed by weaker emission at lower frequencies. In this scenario, the
emission at $t\sim30$~s seen in the main panel is due to material moving
with lower Lorentz factor behind the leading edge of the cocoon shell.
Unfortunately, our simulation did not last long enough to see the
decaying part of the persistent emission. Since the persistent emission is
due to the continued activity of the inner engine, it is a tunable
parameter in our model. A likely assumption is that the duration of
the persistent emission should be similar to the duration of observed
LGRBs.

We have so far shown that, under some assumptions, the large off-axis
angle emission from LGRB progenitors has the properties of SGRBs with
persistent emission components. Our simulation results can also be
used to predict the ratio between long- and short-duration GRBs and
their redshift distributions. To check whether such distributions are
in agreement with observations, we have performed a cosmological
sampling of bursts produced according to this model. First, we have
computed the light-power curves at all off-axis angles and multiplied
them times a radiative efficiency of 50 per cent. At all off-axis
angles, the light-power curves were computed at a fixed radius of
$2.5\times10^{11}$~cm, and therefore all the caveats discussed above
apply to these calculations. We have then generated a sample of 1
million GRBs with random orientation and with a redshift distribution
that follows the star formation rate (we used the star formation rate
SFR3 of Porciani \& Madau 2001) up to redshift 20. Then we have
computed the peak luminosity assuming a cosmology with $h=0.71$,
$\Omega_m=0.27$, and $\Omega_\Lambda=0.73$. Finally, we have applied a
detection threshold of
$F_{\rm{lim}}=8\times10^{-7}$~erg~cm$^{-2}$~s$^{-1}$ for the BATSE
instrument (Nava et al. 2006).

\begin{figure}[!t]
\plotone{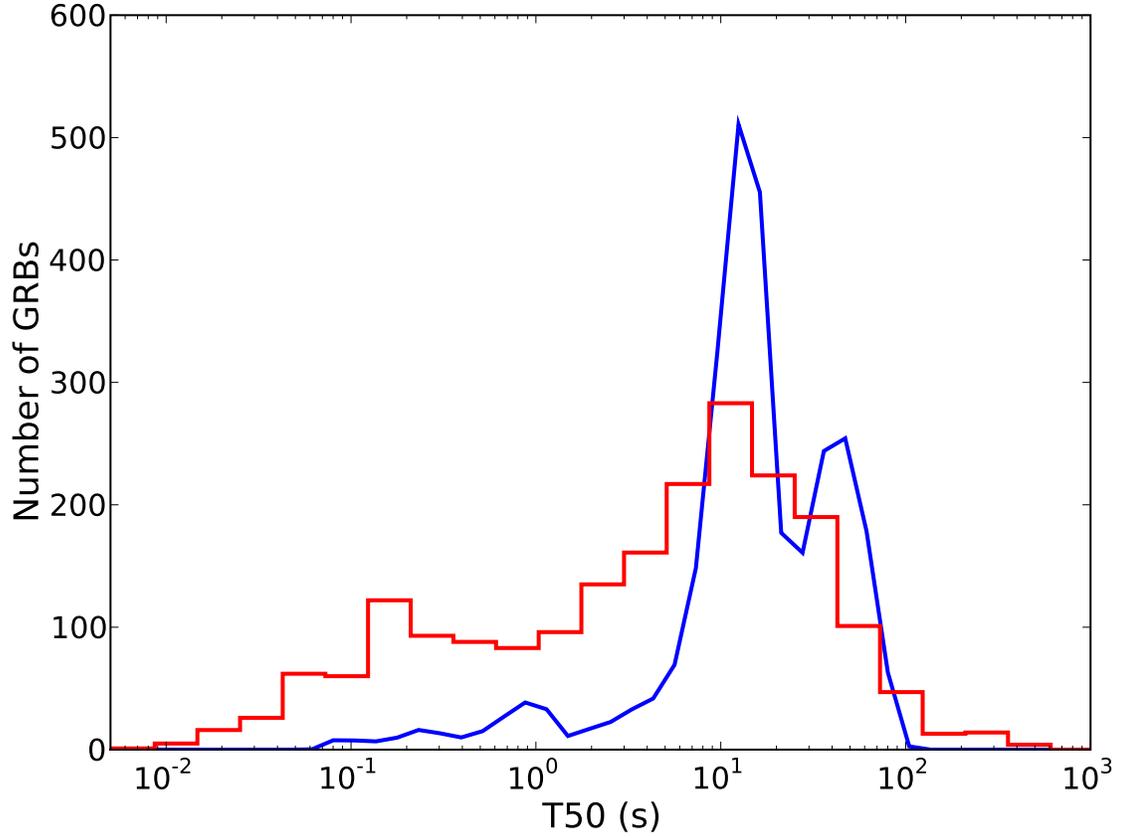}
\caption{{Duration (left) and redshift (right) distributions of GRBs
    from a cosmological sampling of light-power curves from the
    numerical simulation. In the left panel, the red histogram shows
    the $T_{50}$ distribution of 2041 BATSE GRBs, while the blue line
    shows the distribution from the cosmological sampling of our
    model. The synthetic distribution is normalized to have the same
    number of long-duration bursts (those with $T_{50}>2$~s). In the
    right panel, the synthetic redshift distribution of long and short
    GRBs is shown. The predicted average redshift of short GRBs is
    $\langle{z}\rangle=0.8$. The maximum observed redshift is
    $z_{\rm{max}}=1.4$. The true redshift distribution may be broader
    due to the diversity of progenitor and engine properties.}
\label{fig:t50}}
\end{figure}

The left panel of Figure~\ref{fig:t50} shows the observed $T_{50}$
distribution from our cosmological sampling (blue
line)\footnote{$T_{50}$ is the time during which $50\%$ of the fluence
  is detected, measured from the $25^{\rm{th}}$ to the $75^{\rm{th}}$
  percentiles.}. The BATSE duration distribution is overlaid for
comparison. Our synthetic $T_{50}$ distribution result shows clearly
the broad peak of LGRBs, and a tail that extends in the sub-second
durations. The model under-predicts the number of SGRBs, as expected
since some SGRBs are associated with binary mergers\footnote{Only 15\%
  of the BATSE short GRBs are from off-axis collapsars, according to
  this model. The fraction may be higher for HETE-2 and Swift since
  their sensitivity is biased towards lower photon frequencies.}. The
agreement between the synthetic distribution and the BATSE data is
only qualitative as a result of the fact that the synthetic
distribution is based on a single jet/progenitor configuration, while
the real progenitor population is likely characterized by some degree
of diversity.

From the same cosmological sampling of our collapsar model we computed
the redshift distribution of both long- and short-duration GRBs. The
result is reported in the right panel of Figure~\ref{fig:t50}, where
the blue line shows the long GRBs and the red line shows the short
ones.  Again, the results are in qualitative agreement with
observations, with SGRBs detected within a redshift
$z_{\rm{max}}\simeq1.4$ with an average redshift
$\langle{z}\rangle\simeq0.8$. LGRBs have instead a maximum redshift
$z_{\rm{max}}\simeq7$ and an average redshift
$\langle{z}\rangle\simeq2$. Again, the agreement is not exact since
our collapsar simulation cannot reproduce the diversity in progenitor
and engine properties. In addition, the numbers mentioned above depend
on the efficiency that is adopted in the calculation of the
light-power curves. If an efficiency lower than 50\% is selected, the
ratio of long over short GRBs increases, and all the typical redshift
values decrease. In any case, the fact that with a single
progenitor/jet configuration we can obtain such a close qualitative
agreement supports the fact that the orientation-dependent jet-star
interaction is responsible, at least in part, for the diversity of GRB
light curves.

As a final test, we computed the predicted brightness of the afterglow
seen by an observer at large off-axis angles. This is an important test,
since the afterglow at late times will be dominated by the very
energetic core of the jet (Rossi et al. 2002, 2004; Zhang \& Meszaros
2002). We computed afterglow light curves for an off-axis angle
$\theta_o=45\degr$, using the code of Rossi et al. (2004) and assuming
no sideways expansion of the jet material (Zhang \& MacFadyen 2009). We
used equipartition parameters $\epsilon_e=0.1$ and $\epsilon_B=0.01$ and
a slope $p=2.3$ for the non thermal electron population (Panaitescu \&
Kumar 2001). The afterglow is computed for a wind environment (with
$A_\star$=1) and for a uniform interstellar medium (ISM,
$n=1$~cm$^{-3}$) and the flux density is computed assuming a redshift
$z=1$. The parameter $A_\star$ is defined through
$n(r)=3\times10^{35}\,A_\star\,r^{-2}$, where $n(r)$ is the density of
the progenitor wind. $A_\star$ is depends on the mass loss rate and wind
velocity (see Eq. 7 in Panaitescu \& Kumar 2001). The results are
reported in Figure~\ref{fig:aglow}, where a red line shows the R band
afterglow flux density and a blue line shows the 2 keV flux density. For
the wind case, the afterglow has a shallower decays compared to the one
from a uniform fireball. For the uniform case, the afterglow is
initially identical to the one of a spherical fireball, but at
approximately 1 day it flattens, as the bright core of the jet comes
into sight.

\section{Discussion}

\begin{figure}[!t]
\plotone{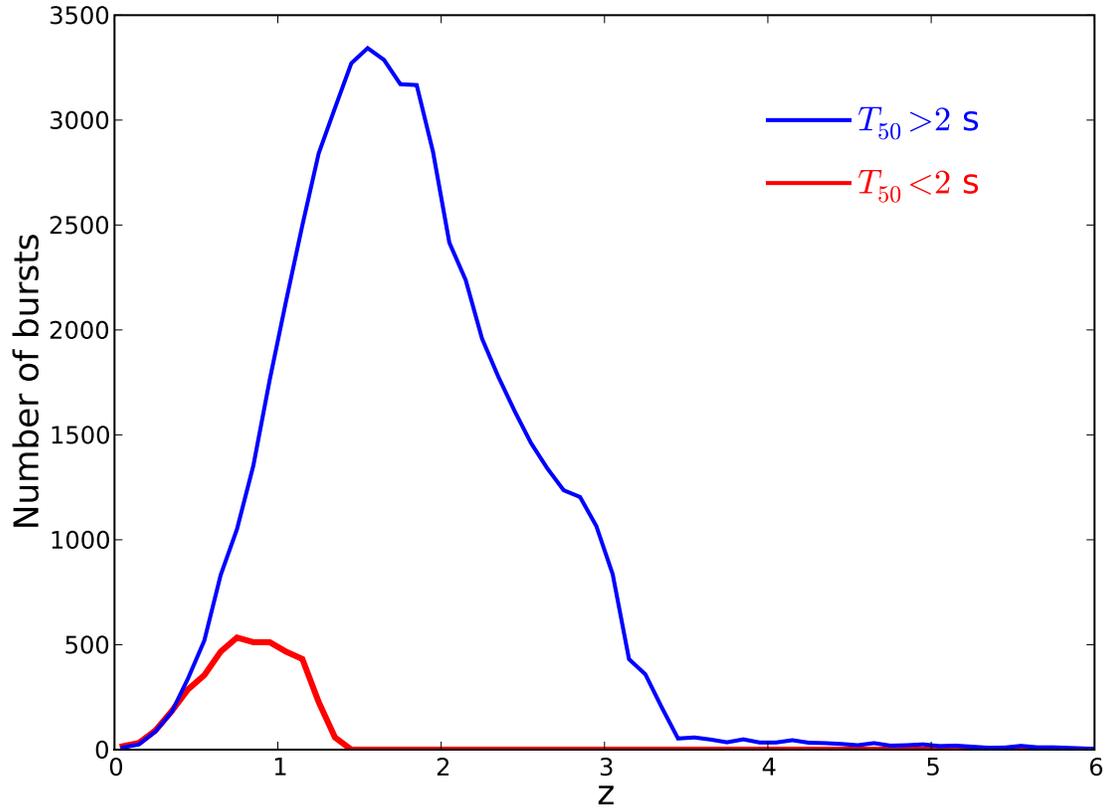}
\caption{{Optical and X-ray afterglow flux density from a burst at $z=1$
observed at a viewing angle $\theta_o=45\degr$. The left panel shows the
afterglow from a wind environment with $A_\star=1$, while the right
panel shows the afterglow from a uniform interstellar medium with $n=1$.
The flattening of the light curves at $t\sim1$~day is due to the
appearance of the bright emission from the on-axis jet and is a
characteristic of this model for short GRBs. The dashed lines show an
analogous calculation for a uniform shell with the same isotropic
equivalent energy of the cocoon shell along the $\theta_o=45\degr$
direction.}
\label{fig:aglow}} 
\end{figure}

We presented the results of numerical simulations of long GRB outflows
from massive progenitor stars within the collapsar scenario. In this
letter, we focused on the outflow at large off-axis angles, showing that
it has a very thin structure and that it can lead to emission with
the characteristics of a SGRB. The presence of slower material behind
the leading edge of the outflow suggests that the emission at lower
frequencies may last longer than the prompt $\gamma$-ray emission, a
characteristic that has been observed in the subclass of SGRBs with
persistent X-ray emission. We propose that at least some SGRBs
with persistent emission are due to collapsars seen at wide angles,
$40\degr$ to $50\degr$ away from the axis along which a LGRB is
released. We check this possibility by computing the number, the
duration distribution, and the redshift distribution of such a short GRB
population, finding that the predictions are in good qualitative
agreement with observed quantities. A better agreement is prevented by
the fact that our simulations do not incorporate the diversity in
progenitor and engine populations of LGRBs. Diversity in the
properties of the GRB progenitor may also cause a small fraction of
short GRBs due to on-axis collapsars with a short-duration engine
(Janiuk \& Proga 2008; Proga et l. 2009).
There are several ways in which the validity of this model can be
checked against observations. The smoking gun of a SGRB from an off-axis
collapsar would be the detection of an accompanying SN explosion in the
late afterglow, as observed in LGRBs (Stanek et al. 2003; Hjorth et al.
2003). Such accompanying SN may however be fairly different from those
accompanying LGRBs, due to the difference in the observing angle. In
particular, SNe associated with off-axis collapsars should be
characterized by slower expansion velocities towards the observer, a
less luminous peak (due to the fact that the $^{56}$Ni is ejected
preferentially along the axis, Maeda et al. 2006), and a longer time to
reach maximum. All these characteristic make the SN detection more
difficult, and could explain the unsuccessful searches of SN components
in several SGRBs\footnote{The other explanation being the fact that
these bursts are not from off-axis collapsars, given their low
redshift.} (GRB~060505 and GRB~060614, Della Valle et al. 2006; Fynbo et
al. 2006; Gal-Yam et al. 2006). Second, if our model is correct, the
host galaxies of some SGRBs should have properties fairly similar to
those of LGRBs. The comparison of host galaxies of SGRBs and LGRBs shows
marked differences (Berger 2009; Fong et al. 2009), especially at low
redshift. At $z\sim1$, the host galaxies of the two populations become
more similar (see, e.g., Figure 3 in Berger 2009), as our model would
predict. Given the small number of galaxies involved, it is premature to
give any firm conclusion, and only further studies with more extensive
samples can give better constraints. Along similar lines, Nysewander et
al. (2009) found that the density of the ambient medium of LGRB and SGRB
afterglows are similar. Finally, SGRBs from off-axis collapsars may be
identified through their afterglow properties, especially if they
explode in a uniform environment (Figure~\ref{fig:aglow}).

The simulation that we have presented is just one case of a big family
of possibilities, with diverse progenitors and diverse engine
properties. The properties of the SGRBs at large off-axis angles are
likely to depend mostly on the luminosity of the jet, with more
energetic events from more luminous jets. The volume of the cocoon
inside the progenitor star affects instead the temperature and Lorentz
factor of the cocoon outflow. Harder SGRBs are therefore expected from
more compact stars. Lacking  a definite radiative model, however, it is
difficult to clearly connect the Lorentz factor of the outflow to a peak
frequency of the spectrum. A definitive measurement of a high Lorentz
factor ($\Gamma>100$) for a SGRB would be inconsistent with this model,
since the cocoon shell can hardly have Lorentz factors $\Gamma>20$.
Finally, short bursts from off-axis collapsars should have little
internal variability, unless some localized dissipation process, such as
magnetic reconnection (Lyutikov \& Blandford 2003) of relativistic
turbulence (Narayan \& Kumar 2009), provides the internal energy for the
radiation of photons.

In conclusion, we would like to stress that what we presented is a
set of necessary conditions for the production of SGRBs. These are by no
means sufficient and at least two important assumptions have to be made
to successfully predict SGRBs from our simulation. First, we have to
assume that all the radiation is released at a fairly constant and small
radius. Should this fail, the light curve would be smeared on a longer
time and the burst would not be classified as a SGRB. Second, we have to
postulate that the unknown radiation mechanism is able to produce
photons in the MeV range, even if it expands at a much slower Lorentz
factor compared to the on-axis jet. Should this fail, the cocoon
emission would likely produce an x-ray transient rather than a SGRB.

\acknowledgements We would like to thank the anonymous referee for a
careful and constructive review and Giancarlo Ghirlanda for his help
with BATSE detection thresholds and Edo Berger, Massimo della Valle,
Asaf Pe'er, Daniel Proga, and Bing Zhang for useful comments. The
software used in this work was in part developed by the DOE-supported
ASC / Alliance Center for Astrophysical Thermonuclear Flashes at the
University of Chicago. This work was supported in part by NASA ATP grant
NNG06GI06G and Swift GI program NNX06AB69G (MB) and NNX08BA92G (DL). We
thank NASA NAS for the generous allocations of computing time.

\newpage

\end{document}